# Energy Control Strategy to Enhance AC Fault Ride-Through in Offshore Wind MMC-HVDC Systems

Dileep Kumar
Department of Electrical and Computer Engineering
University of Houston
Texas, United States
dkumar20@cougarnet.uh.edu

Wajiha Shireen
Department of Electrical and Computer Engineering
University of Houston
Texas, United States
tech139@central.uh.edu

*Abstract*—Modular Multilevel Converter-based High Voltage Direct Current (MMC-HVDC) system is a promising technology for integration of offshore wind farms (OWFs). However, onshore AC faults on MMC-HVDC reduce the power transfer capability of onshore converter station, leading to surplus power accumulation in HVDC link. This surplus power causes a rapid rise in DC-link voltage and may hinder safe operation of OWFs. To address such a situation, this paper presents an AC fault ride-through scheme that combines the storage of surplus power in MMC submodule (SM) capacitors and dissipation of residual power in an energy dissipation device (EDD). The proposed energy control facilitates use of half-bridge MMC SMs with low-capacitance, with their storage capacity leveraged to share the surplus power during faults, with a lower-rated EDD. The proposed scheme is tested on a ±320kV/420MW MMC-HVDC system. The results show that proposed control scheme effectively maintains DC link voltages, ensuring connection of OWFs.

## I. Introduction

Offshore wind power has recently gained momentum because of its inherent advantages such as strong stability, higher wind potential and no occupation of onshore land resources. Offshore wind power is one of the large capacity renewable energy sources with the highest rate of growth [1-4]. With the increasing need to integrate far-off offshore wind power into onshore grids, AC transmission has become uneconomical due to losses in skin effect and higher cost of cables [5].

HVDC transmission, conversely, is the most economical solution for delivering large offshore power, offering lower losses, interconnection of asynchronous systems, and no skin effect. A voltage source converter-based HVDC (VSC-HVDC), particularly MMC-HVDC is regarded as the appropriate choice for remote offshore wind power transmission due to its decoupled active and reactive power control, enhanced power system stability, high transmission capacity for undersea power links, and power oscillation damping capability. The first MMC-HVDC connection Trans Bay Cable project (±200 kV/400 MW), which links Pittsburgh, Pennsylvania, and San Francisco, California was commissioned in 2010. A point-to-point MMC-HVDC system connecting OWFs to onshore AC grid is shown in Fig.1, with MMC1 serving as the receiving-end converter (REC) and MMC2 as the sending-end converter (SEC). When a short-circuit fault occurs at the grid side of REC, the upper limit of REC converter's output AC power to the grid is greatly reduced. Whereas the power from OWF stays steady and continues to charge the capacitance of HVDC transmission system, causing DC link over-voltage, which may result in insulation breakdown. In severe cases, such as a three-phase-to-ground (LLL-G) short circuit fault at the onshore ac grid, DC link voltage may rise over its safe limit in as little as 30 to 50 milliseconds. Furthermore, this puts stress on equipment and may cause activation of protection scheme, leading to disconnection of the onshore converter station from the AC grid.

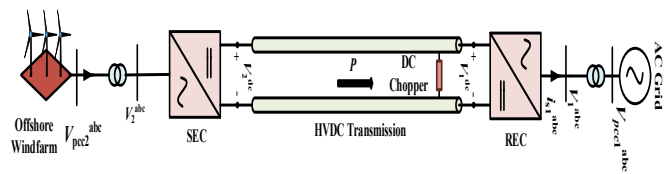

Figure 1. MMC-HVDC Systems connecting OWFs and onshore AC Grid

Traditional methods to solve DC link overvoltage are: a) surplus power dissipation in AC/DC choppers or EDDs, and b) direct reduction of wind power. Power dissipation in EDD during grid side AC faults is simple and requires no reconfiguration in control. More than ten offshore projects have adopted choppers [6], including Borwin6 project in Germany and Rudong project in China [7]. Although the technologies for power dissipation are quite simple, they come up with capital costs and bring logistic and shipping challenges. For the active power reduction through frequency-up or voltage-down methods, rapid curtailment of wind turbines generation is challenging. Due to inertia, reduction rate of wind turbine power is slow, typically 100ms to seconds to bring rated power to zero. While the DC link voltages rise quickly within tens of milliseconds. To reduce the cost of EDD, [8] proposed a DC chopper based on thyristor-LC (TLC-DCC). Still, the cost of installing chopper remains a major challenge. [9] proposed a communication-based wind turbine power reduction method during faults. The proposed scheme effectively controls the fault situation; however, communication links have inherent communication delays and may cause reliability issues. In the offshore wind VSC-HVDC setup of Fig. 1, SEC uses voltage/frequency (V/f) control. When a short-circuit fault happens on ac grid side, a signal is sent to the SEC to reduce the wind farm's output voltage, which has a communication delay of about 30–50ms. In addition to this, wind turbines take another 100ms to make their output zero. Therefore, during this period of FRT, surplus energy must be managed.

The current trend is to store the surplus power in MMC submodule (SM) capacitors [10,11], however, literature review shows that this requires larger SM capacitance which will potentially scale up the size and weight of the converters. This will pose significant economic and technical challenges in the construction and installation process of offshore platforms. Taking an example of BorWin3-Offshore platform weighing 30,000-ton came with significant costs during development and shipping. Generally, in defining the overall mass of an offshore converter station, the sizing of the SM capacitor in an MMC plays a crucial role, with SM capacitors accounting for approximately 60% of the total mass. Also, these methods to store surplus power in MMC SM capacitors only solve scenarios of single-line-to-ground (SL-G) short circuit faults or partial drop of AC voltages and may not handle the worst fault scenarios, such as LLL-G faults [12]. [7] uses the power-sharing concept between SM capacitors and a series-connected energy diverting converter (SC-EDC), however, the use of full bridge MMC (FB-MMC) topology will make the offshore converters bulky and implementation requires a communication link.

Considering the challenges mentioned above, this paper proposes a FRT scheme, which combines a lower-rated DC chopper and storage capacity of half-bridge MMC SM capacitors to store excessive power. The idea is to tap the half-bridge SM capacitors capacity. When SM capacitors reach the maximum capacity, the residual power is dissipated in an EDD. This reduces the EDD size while also utilizing the energy storage margin of SM capacitors. Furthermore, considering the weight constraints of offshore platform, the EDD is placed on the onshore side. The major contributions of this paper are:

1) An AC FRT scheme based on combination of surplus power storage in MMC SM capacitors and residual power dissipation in an EDD is proposed.
2) A low-rated EDD is employed, which reduces the overall capital cost associated with riding-through faults. Its operation during SL-G and LLL-G faults is analyzed
3) MMC SMs with low capacitance are utilized, with their storage capacity leveraged to absorb the surplus power during faults.

The rest of this paper is organized as follows: Section II proposes an MMC-HVDC control scheme. In Section III, the effectiveness of the proposed method is verified in FRT. Finally, Section IV concludes the work.

## II. PROPOSED FRT METHOD

Fig. 2 shows the half bridge topology of three-phase MMC. Each phase $j$ ($j= a, b, c$) has upper and lower arms with series-connected submodules SMs (1 to $N$).

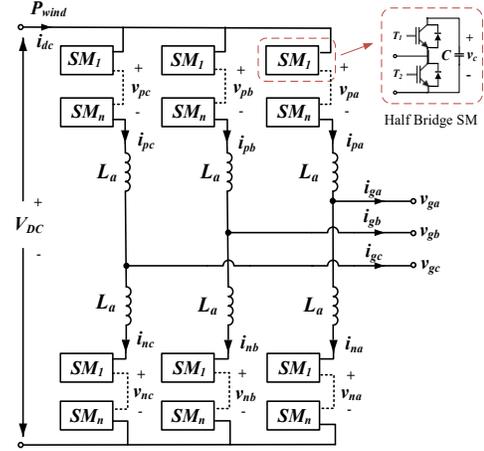

Figure 2. Topology of MMC

The output voltages of each arm of MMC have three components: DC component $(V_{dc})$, AC component $(v_{ac})$ and circulating component $(\Delta v_{pn})$.

$$\begin{cases} v_p^j = \frac{1}{2}V_{dc} - v_{ac}^j - \Delta v_p^j \\ v_n^j = \frac{1}{2}V_{dc} + v_{ac}^j - \Delta v_n^j \\ i_{gj} = i_{pj} - i_{nj} \end{cases} \quad (1)$$

Where $i_{gj}$ is AC grid current, $i_{pj}$ is current in upper arm and $i_{pj}$ is current in lower arm. From (1), it can be concluded that different parts of arm voltages can be controlled independently to control the arm currents in MMC.

### A. SM Capacitor for Surplus Power Storage

SM capacitors are the primary energy storage elements of MMC converters. As seen from Fig. 2, MMC has numerous SMs with parallel connected capacitors. By proper control, surplus energy can be stored in SM capacitors. According to capacitor selection principle in engineering applications, the capacitor voltage can go 1.5 times the rated voltage [11], providing a margin to store the energy during fault. Under normal conditions, the energy in both SEC (offshore) and REC (onshore) is given by:

$$E_{MMC} = 3N \times (CV_0^2) \quad (2)$$

Where $E_{MMC}$ is MMC energy, $N$ is number of submodules in an arm, $C$ is SM capacitance and, $V_0$ is the capacitor voltage.

On the occurrence of a grid side ac fault, the surplus power is expressed as:

$$P_s = P_{wind} - V_{pcc1} I_{ac} \quad (3)$$

$$V_1 = \sqrt{\frac{\int P_s dt}{3NC} + V_0^2} \quad (4)$$

Where $P_{wind}$ is power supplied from the OWF, $V_{pcc1}$ is voltage at point of common coupling (PCC) on the onshore side after the fault at onshore AC side of grid, and $I_{ac}$ is the current. The fault causes DC link voltages to go above the threshold (i-e.1.1 P.U.), and the SM capacitors of MMC converters are allowed to absorb the surplus energy which can be expressed as:

$$\Delta E_{MMC} = 3NC \times (V_1^2 - V_0^2) \quad (5)$$

Where $\Delta E_{MMC}$ is change in MMC energy, N is number of submodules, C is SM capacitance and, $V_0$ and $V_1$ are pre and post fault capacitor voltages. Considering the worst-case scenario, i.e., an LLL-G fault, the SM capacitors alone cannot store the surplus power, because there is a transient voltage spike of hundreds of volts higher than capacitor voltage when IGBT is turned off which may damage the IGBT [12]. Therefore, [12] proposes a voltage-boosting circuit for MMC SM. Considering the limitations of storage capacity of half bridge SM driven by sudden voltage spike, this paper uses an EDD. Once the SMs' overvoltage capacity is fully utilized, the residual power is dissipated through a low-rated EDD:

$$\Delta E_{EDD} + \Delta E_{MMC} = \int_{t_0}^{t_1} P_s dt \quad (6)$$

Fig. 3 shows the flowchart of the proposed scheme. The complete fault ride-through scheme is divided into two stages: surplus energy storage in half-bridge SM capacitors and residual energy dissipation in the EDD. On the occurrence of an onshore AC fault, the surplus power is calculated using (3), based on which the DC modulation signal ($M_{dc}$) is generated to dynamically update the upper and lower arm modulation signals ($m_u$ and $m_l$). This modulation adjustment enables the half-bridge SM capacitors to absorb the surplus energy. Once the SM capacitors reach their maximum energy capacity and if the fault remains uncleared, the residual surplus energy is dissipated in the EDD, ensuring DC link voltage stability and system operation.

Taking the parameters used in this study: C=3000 uF, N=76 and $V_0$=8.42kV and using (2) the nominal energy in individual MMC half-bridge SM capacitors ($E_{MMC}$) is 48.5 MJ (1.0 P.U.). However, during the fault, capacitor voltages change, leading to change in total MMC energy. Using (5) energy absorption in onshore converters ($E_{REC}$) during LLL-G increases to 80.025 (1.65 P.U.) MJ while $E_{SEC}$ is 58.2MJ (1.2 P.U.). While the residual energy $\Delta E_{EDD}$ of approximately 26.05 MJ, is dissipated in the form of heat.

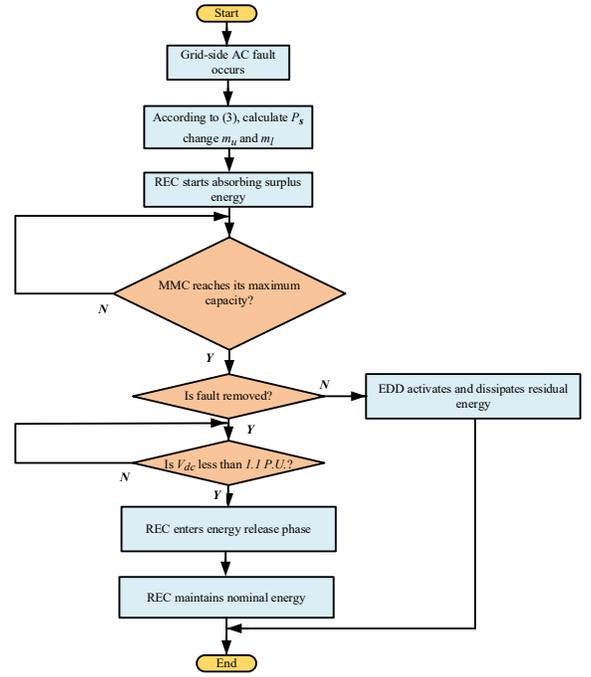

Figure 3. Flowchart of the proposed method

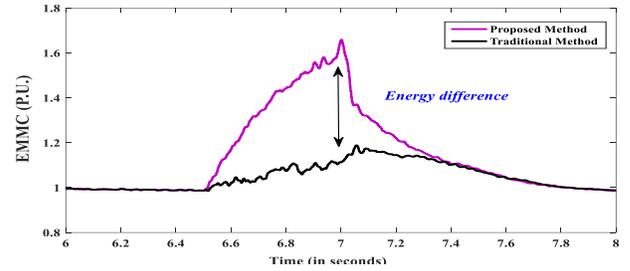

Figure 4. Energy Difference in SM Capacitors during LLL-G at t= 6.5s

### B. Implementation of an EDD

The worst-case scenario occurs when an LLL-G fault takes place on the onshore AC side, significantly reducing power export, while the wind turbines operate at their maximum power output. Under such a scenario, the EDD in a point-to-point connection must be rated equal to the full power output of the wind turbines. The following formula represents the active power delivered from the converter to the grid [10]:

$$P_{wg} = \frac{V_w V_g \sin\emptyset}{X} \quad (7)$$

Where, $P_{wg}$ is the power transferred to the grid, $V_w$ is voltage at sending end, $X$ is reactance of line and $V_g$ represents the receiving-end voltage. During an LLL-G fault, $V_g$ drops to zero. Consequently, according to (7), onshore converter power transfer to the grid becomes zero, while the offshore wind turbines maintain their pre-fault power, causing power mismatch and DC link voltage rise. If all this wind power were to be dissipated solely by the EDD by circulating current through a resistor, it would significantly increase the overall size and cost of the chopper. In this study, the EDD is activated when DC link voltage surpasses 1.06 P.U. and dissipates energy proportionally to the voltage deviation. Two key design parameters to be considered are the instantaneous power dissipation $P$(in watts) and the total energy dissipation $E$(in joules). The latter is determined by integrating the dissipated power over the time interval $T$. The energy dissipated in the EDD is expressed by:

$$E = \int_{t_0}^{t_1} P_s dt \qquad (8)$$

*C. Control of MMC*

In MMC-HVDC systems, the upper-level control is typically divided into outer-loop and inner-loop control. For OWF MMC-point-to-point MMC-HVDC, the REC's outer-loop control manages either the DC-link voltage or AC voltage. The output of the outer loop serves as a reference for the inner current-control loop. While, the SEC (i.e., MMC-2) regulates the AC voltage and frequency, enabling operation in islanded mode. To decouple DC link voltage and SM capacitor voltage, modulation dimension is increased to two: AC modulation ($m_{ac}$) and DC modulation ($M_{dc}$):

$$\begin{cases} m_u = M_{dc} + m_{ac} \\ m_l = M_{dc} + m_{ac} \end{cases} \qquad (9)$$

When onshore AC fault occurs, the surplus power is calculated using (3). Depending on the severity of fault, DC modulation signal ($M_{dc}$) combined with the AC modulation signal ($m_{ac}$), updates the upper and lower arm modulation signals ($m_u$ and $m_l$) It is important to note that to avoid a communication signal transfer from onshore to offshore converter during fault and to prevent delays, only the onshore converter is allowed to store surplus energy; therefore, DC modulation is applied only to the onshore converter station. Fig. 4 compares the difference in capacitors' energy of REC during and after LLL-G fault with and without DC modulation. Fig. 5 shows the complete control block. Under normal operating conditions, both the onshore and offshore converters maintain their nominal energy levels, and the EDD remains idle. However, during an onshore AC fault, $M_{dc}$ is generated at the onshore MMC, enabling its half-bridge SMs to absorb the surplus energy. Once the SM capacitors reach their maximum capacity, the EDD is activated to dissipate residual energy.

## III. RESULTS DISCUSSION

The effectiveness of the proposed scheme has been tested on a point-to-point MMC-HVDC system in PSCAD/EMTDC for both AC LLL-G and SL-G faults. Table I shows the system parameters. At this point, it is important to note that MMC-1 is connected to the onshore AC grid while the MMC-2 is connected to the OWFs.

*A. LLL-G Fault*

In this scenario, an LLL-G fault is introduced at t = 6.5 seconds for the duration of 500ms. Fig. 6 shows the results when the fault occurs on the onshore AC side of MMC-1. As seen in Fig.6, upon the occurrence of the fault, the DC link voltage rises but remains within the safe limit.

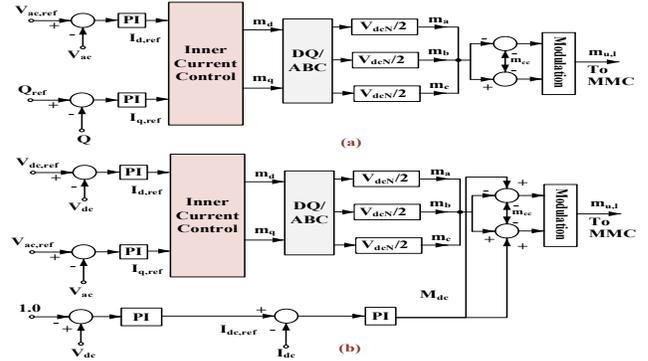

Figure 5. Overview of MMC Control: (a) Control of SEC (offshore MMC) (b) Control of REC (onshore MMC)

The voltage at the onshore MMC-1 station drops to zero. Since power is the product of current and voltage, the power at MMC-1 also becomes zero, while the power supplied from the OWFs remains the same. The SM capacitors in REC start absorbing the surplus energy. When their energy reaches 1.65 P.U., the EDD gets activated and the residual energy of approximately 26.05 MJ is dissipated in form of heat. When the fault is cleared, the system returns to its pre-fault condition, maintaining the stable connection.

*B. SL-G Fault*

Fig. 7 shows the results for the system when a SL-G fault occurs on the onshore AC side of MMC-1. The fault occurs at t=6.5 seconds for the duration of 120ms. As soon as the fault occurs, one phase of the three-phase voltages goes to zero, while the other two phases remain unaffected, as shown in Fig. 7. A small rise in DC link voltage is observed. In a SL-G fault, 2/3 of the total power is reduced, which is observed at onshore MMC-1, while the power supplied from the OWFs remains unaffected. When the fault is cleared, the system returns to its normal condition. SM capacitors in REC enter energy absorption phase and store maximum energy of 1.32 P.U. while around 6 MJ is dissipated in EDD in form of heat. Fig. 7 (e) and (f) shows the energy variation of MMC SM capacitors and dissipation in EDD respectively, during and after the fault. The results demonstrate that the proposed FRT scheme effectively maintains the DC link voltage by absorbing excess power from the OWFs in the MMC SM capacitors. This approach helps reduce the overall cost associated with EDD installation, as well as shipping and logistical challenges.

TABLE I. SYSTEM PARAMETERS

| Parameter | Value | Unit |
| --- | --- | --- |
| Rated Active Power $P_{wind}$ | 420 | MW |
| DC-link Voltage $V_{dc}$ | ±320 | kV |
| Number of SMs per Arm $N$ | 76 | - |
| Nominal Frequency $f$ | 50 | Hz |
| MMC SM Capacitance $C_{SM}$ | 3000 | uF |
| MMC SM rated voltage | 8.42 | kV |

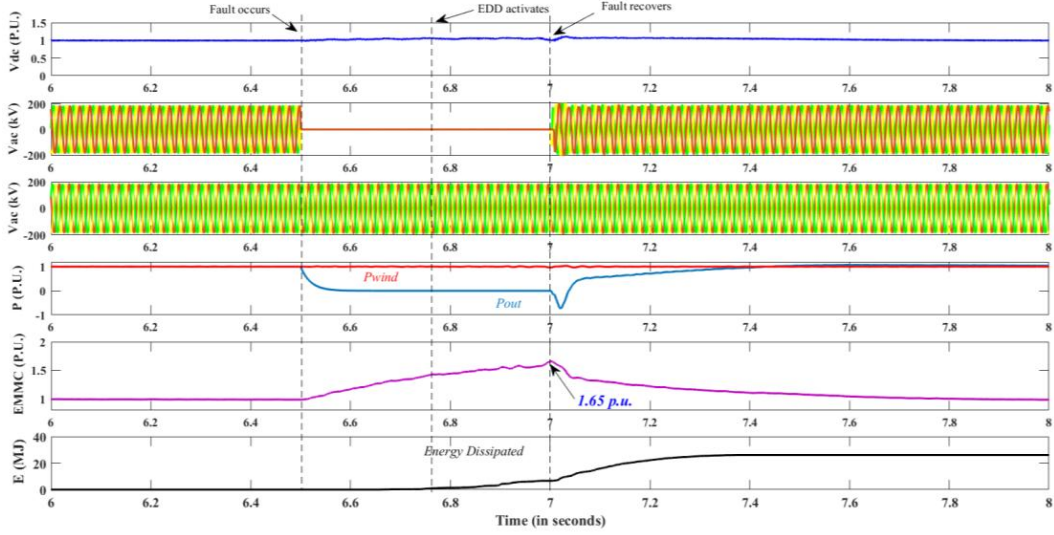

Figure 6. LLL-G fault: (a) DC-link voltage (b) Grid-side AC voltages (c) Wind-side AC voltages (d) Active Power (e) MMC Energy (f) Energy Dissipated in EDD

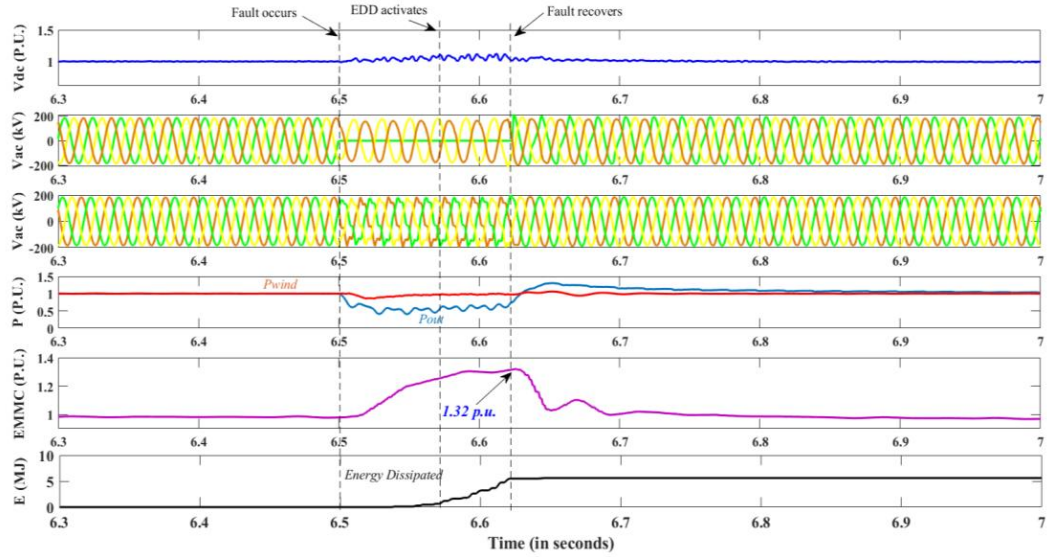

Figure 7. SL-G fault: (a) DC-link voltage (b) Grid-side AC voltages (c) Wind-side AC voltages (d) Active Power (e) MMC Energy (f) Energy Dissipated in EDD

## IV. Conclusion

In this paper, an energy control scheme for FRT of MMC-HVDC systems is proposed. The scheme combines MMC SM capacitors and an EDD to store and dissipate excess energy, thereby maintaining stable DC link voltages. The proposed approach has been tested in PSCAD/EMTDC under both LLL-G and S-LG fault scenarios. In both cases, the DC link voltages remain within safe limits while maintaining the connection to the onshore AC grid. After fault clearance, the system returns to its pre-fault operating conditions. Future work will focus on leveraging the energy storage capacity of the MMC SM capacitors and reducing the size and capital cost of the EDD.